\documentclass[%
reprint,
groupedaddress,
amsmath,amssymb,
aps,
prx,
]{revtex4-2}

\renewcommand{\thesection}{\arabic{section}}
\renewcommand{\thesubsection}{\thesection.\arabic{subsection}}
\renewcommand{\thesubsubsection}{\thesubsection.\arabic{subsubsection}}

\usepackage{amsmath}
\usepackage{graphicx}
\usepackage[breaklinks=true,colorlinks=true,linkcolor=blue,urlcolor=blue,citecolor=blue]{hyperref}
\usepackage{units}
\usepackage{upgreek}

\begin{document}

\title{Optical antennas driven by quantum tunneling: a key issues review}

%

\author{Markus Parzefall}
\affiliation{Photonics Laboratory, ETH Z\"urich, 8093 Z\"urich, Switzerland}

\author{Lukas Novotny}
\affiliation{Photonics Laboratory, ETH Z\"urich, 8093 Z\"urich, Switzerland}

\begin{abstract}
Analogous to radio- and microwave antennas, optical nanoantennas are devices that receive and emit radiation at optical frequencies. Until recently, the realization of \emph{electrically} driven optical antennas was an outstanding challenge in nanophotonics. In this review we discuss and analyze recent reports in which quantum tunneling---specifically inelastic electron tunneling---is harnessed as a means to convert electrical energy into photons, mediated by optical antennas. To aid this analysis we introduce the fundamentals of optical antennas and inelastic electron tunneling. Our discussion is focused on recent progress in the field and on future directions and opportunities.
\end{abstract}


\maketitle


\tableofcontents


\section{Introduction}



An antenna in the `classical' context is a device that converts electronic signals to electromagnetic waves and vice versa. The 
wide applicability of this principle becomes clear when considering the frequency range over which it is applied. 
Antennas are commonly used from frequencies as low as several \unit{kHz} to frequencies as high as hundreds of \unit{GHz}, spanning approximately eight decades. 


The size of antennas is on the order of the wavelength $\lambda$, which scales inversely proportional to the operating frequency $f$. In free space, $\lambda = c / f$ where $c$ is the propagation speed, i.e. the speed of light. 
The wavelength of electromagnetic radiation emitted by `classical antennas' consequently 
ranges between approximately one \unit{mm} and several \unit{km}. 


\subsection{Optical antennas -- antennas for light}
\label{sec:lightantennas}

When we think of an antenna we much rather associate it with an \textit{electronic component} as opposed to an element which we use to manipulate light. This is mostly due to historical reasons. The motivation which led to the invention of the antenna concept originated out of the need to transmit signals over large distances. Light on the other hand has historically been controlled in its freely propagating form by lenses and mirrors, governed by the laws of reflection, refraction and diffraction. 

The optical antenna concept emerged out of the field of microscopy with the motivation to break one of its fundamental limitations, the diffraction limit. 
It limits the resolution of conventional optical microscopes working with visible light to several hundreds of \unit{nm}. However it only applies to propagating waves. It was already realized at the beginning of the 20th century by Edward Hutchinson Synge that one could use a nanoscopic particle, smaller than the wavelength of light, in order to localize the light field and thereby enhance resolution \cite{novotny07b}. The reason for this is that the nanoparticle acts as a tiny resonator for light, momentarily storing energy in its vicinity, i.e. the nearfield, before it is reradiated into the farfield. It took 60 years until Synge's proposal was realized experimentally by Ulrich Fischer and Dieter Pohl~\cite{fischer89}, in an apparatus which is now known as the nearfield microscope~\cite{betzig92a,pohl93a,courjon94}.

The probe of a nearfield microscope essentially solves an antenna problem~\cite{pohl99} in that it ``efficiently converts free-propagating optical radiation to localized energy, and vice versa''~\cite{bharadwaj09a}, rendering it the first realization and utilization of an \emph{optical antenna}. Optical antennas ~\cite{bharadwaj09a,novotny11a,biagioni12a,novotny06b,giannini11,agio13a} and more generally \emph{plasmonics} \cite{ozbay06,maier07,schuller10,gramotnev10,stockman11,kauranen12,zhang13b} today are active fields of research.

Optical antennas can be likened to their macroscopic radio- and microwave counterparts in many aspects. Design principles which have been originally developed for antennas in other frequency domains have been successfully applied to optical antennas \cite{muehlschlegel05,schuck05,taminiau07,curto10,kosako10}. Engineering geometry and arrangement of antenna constituents allows for the tuning of their operating frequencies as well as the determination of their spatial radiation characteristics---ranging from omni- to unidirectionality \cite{balanis05}. Besides a modified scaling behavior due to the material response of metals at optical frequencies \cite{novotny07a}, the main characteristic in which they differ thus far is their mode of operation. 



\begin{figure}
	\centering
	\includegraphics{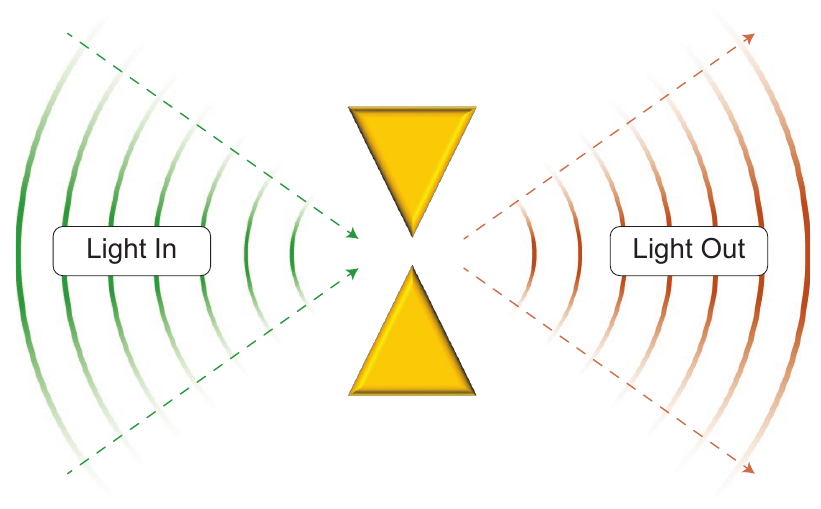}
	\caption{Operating principle of optical antennas that are driven by light.}
	\label{fig:light}       
\end{figure}

Optical antennas have been historically operated on a `light-in/light-out' basis, i.e. the antenna receives radiation from the far-field as well as emits radiation back into the far-field \cite{novotny11a}. This operating principle is schematically depicted in Fig.~\ref{fig:light}. In this context, optical antennas are used for example in spectroscopy in order to strongly localize light fields which facilitates improved coupling between light and nanoscopic entities such as single molecules \cite{frey04,anger06a,muskens07,taminiau08b,kinkhabwala09}, quantum dots \cite{farahani05,curto10}, nitrogen vacancy centers \cite{schietinger09} and carbon nanotubes \cite{hartschuh03a,hayazawa03}. Similarly, strongly enhanced fields are used to enhance weak nonlinear optical processes \cite{palomba09a,kauranen12}. This paradigm of all optical operation has recently started to shift towards the integration of optical antennas in optoelectronic devices for photovoltaics \cite{atwater10}, photodetection \cite{tang08,knight11}  and surface plasmon excitation \cite{walters10,fan12,huang14}.


\subsection{Electrically driven optical antennas}
%

\begin{figure}
	\centering
	\includegraphics{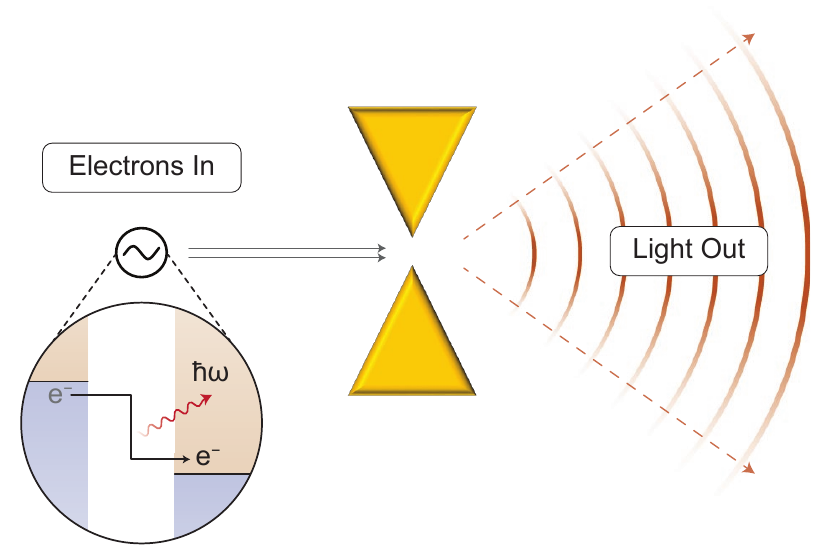}
	\caption{Operating principle of optical antennas that are driven by quantum tunneling. The inset shows a schematic of the inelastic electron tunneling process.}
	\label{fig:tunneling}       
\end{figure}


Antennas were initially invented to solve a communication problem. The same could be true for optical antennas. Electrically driven optical antennas could serve as highly integrable, on-chip, nanoscopic transducers, facilitating the conversion of electrical to optical signals. Furthermore, they could be used as nanoscopic sources of light for spectroscopy and sensing. The question is, how do we drive them electrically? Sources which are traditionally used to feed antennas are limited to frequencies of approximately \unit[300]{GHz} \cite{sirtori02}. At the other end of the spectrum, electromagnetic radiation in the infrared and visible spectral regions are generated by semiconductor LEDs and lasers. 


\begin{figure*}
	\centering
	\includegraphics{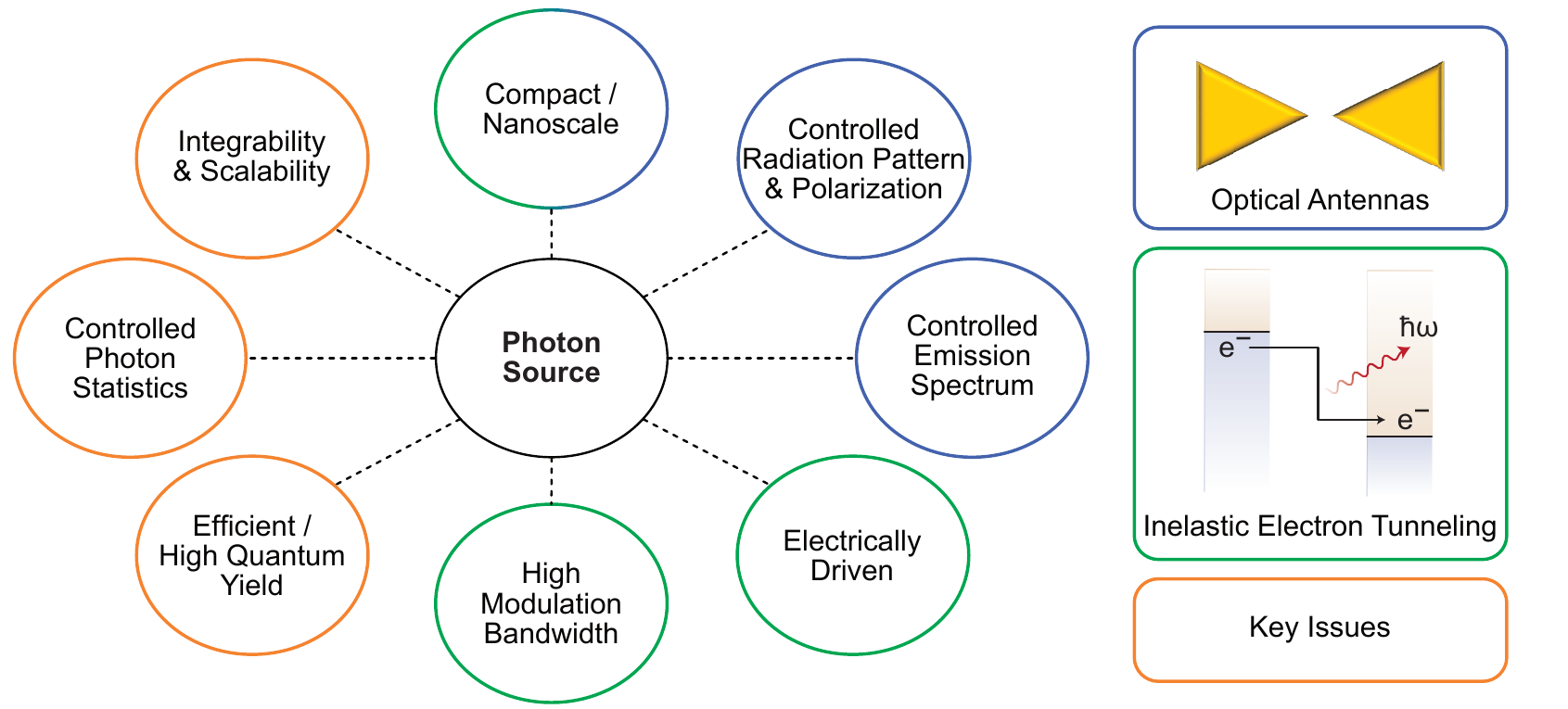}
	\caption{Key prospects and issues of optical antennas driven by quantum tunneling. Optical antennas provide a means to control emission pattern, polarization and spectrum. IET provides the electrical drive and a high modulation bandwidth. The current key issues are low quantum efficiencies, a lack of control over the photon emission statistics as well as moderate device integrability and scalability.}
	\label{fig:properties}       
\end{figure*}


An alternative mechanism for the generation of light was discovered more than forty years ago by Lambe \& McCarthy, reporting the emission of light from a metal-insulator-metal (MIM) tunnel junction~\cite{lambe76}. The physical mechanism they proposed is depicted in the inset of Fig.~\ref{fig:tunneling}.  Two metallic electrodes are separated by an insulating barrier. The barrier is thin enough to allow for electron tunneling between the electrodes. Applying a voltage $V_{\rm b}$ supplies electrons in one electrode with excess potential energy $|e V_{\rm b}|$, where $e$ is the elementary charge. The majority of tunneling processes is elastic, i.e. electrons maintain their energy when traversing the barrier. Alternatively, electrons may also tunnel inelastically (cf.~Fig.~\ref{fig:tunneling}) by coupling to an optical mode of energy $\hbar \omega$, where $\hbar$ is the reduced Planck constant and $\omega = 2 \pi f$ is the angular frequency. Initially, light emission from IET was studied in macroscopic MIM tunneling devices \cite{mills82,szentirmay91}. About ten years later, light emission from a scanning tunneling microscope (STM) was reported by Gimzewski et al.  \cite{coombs88,gimzewski88,gimzewski89}. Since then, photon emission mapping with molecular \cite{berndt93b,cavar05} and even atomic resolution \cite{berndt95,thirstrup99,nazin03,hoffmann04}, spectroscopic analysis of molecular vibrations \cite{qiu03,dong04,dong10,chen10}, imaging of electronic wavefunctions and molecular orbitals \cite{schull08,chen09b,lutz13,yu18} as well as visualization of intermolecular coupling \cite{zhang16} have been demonstrated. In related experiments, STM excitation of localized resonances of nanostructures has been studied \cite{silly00,myrach11,lemoal13,lemoal16}. For more information on this active field of research, see Refs.~\cite{berndt98,sakurai05,rossel10,kuhnke17}.  

\subsection{Scope and overview}

In recent years, IET has gained attention in the optics community as a potential electronic drive for optical antennas as schematically depicted in Fig.~\ref{fig:tunneling}. The scope of this review is to analyze the progress that has been made in this regard. In particular our aim is to point out the key prospects of optical antennas driven by quantum tunneling that have already been demonstrated and continue to motivate further research. More importantly, we will discuss the key issues that need to be addressed in current and future research. Figure~\ref{fig:properties} displays a selection of desirable properties of a photon/light source. The strong prospects of optical antennas driven by quantum tunneling are that the antenna improves the electron-to-photon conversion efficiency and its design determines radiation pattern, polarization and emission spectrum. They furthermore feature an inherently nanoscopic footprint. 

On the other hand, the IET process constitutes an electronic  drive for optical modes. It is thought to be ultrafast as the ultimate limiting time-scale of mode excitation is given by the electron tunneling time, a femtosecond scale process \cite{landauer94,shafir12,fevrier18}. Moreover, a tunnel junction is arguably the smallest possible light source. However, the primary key issue of the IET process is that electron-to-photon conversion efficiencies, i.e. the quantum yield of the device, until recently remained below $\sim 10^{-3}$. Another key issue that needs to be addressed is to find implementations that are not only efficient but also scalable. Lastly, the photon statistics of light emitted due to IET has only recently found some attention with the question of whether it can be controlled remaining to be answered.

For completeness we would like to emphasize that---especially in the high conductivity regime---alternative mechanisms such as hot-electron emission \cite{buret15,cazier16,dasgupta18} and inelastic processes involving more than one electron \cite{schull09,schneider13,xu14,kaasbjerg15,peters17} play an important role in the light emission from tunnel junctions . Recently it was also suggested that a ballistic nanoconstriction may yield a higher emission efficiency than the more widely studied tunneling gaps \cite{uskov17}. 

This review is structured as follows. We will first give brief introductions to both the fundamentals of optical antennas and of inelastic electron tunneling in Sect.~\ref{sec:optant} and Sect.~\ref{sec:IET}, respectively. Next, we will analyze recent reports of experimental realizations of antenna-coupled tunnel junctions with a focus on the key issues and prospects displayed in Fig.~\ref{fig:properties} in Sect.~\ref{sec:expprog}. Finally, some future directions and potential strategies are discussed in Sect.~\ref{sec:outlook}.

\section{Fundamentals of optical antennas}
\label{sec:optant}


Optical antennas\index{optical antenna} are transducers between localized and freely propagating electromagnetic energy at the nanoscale. More specifically, metallic nanostructures---depending on their geometry and dielectric environment---are able to resonantly enhance the electromagnetic energy in nanoscopic volumes, cf. Fig.~\ref{fig:light}. In the context of this review the emission properties of optical antennas are of particular importance and will be briefly introduced. We will further draw important connections between the power radiated by a classical point dipole and the spontaneous emission rate of a quantum mechanical two-level system. Further information on the topic and more detailed introductions can be found in Refs.~\cite{novotny06b,novotny11a,bharadwaj09a,biagioni12a,agio13a}.

\subsection{Antennas for emission enhancement}

\begin{figure}
	\centering
	\includegraphics{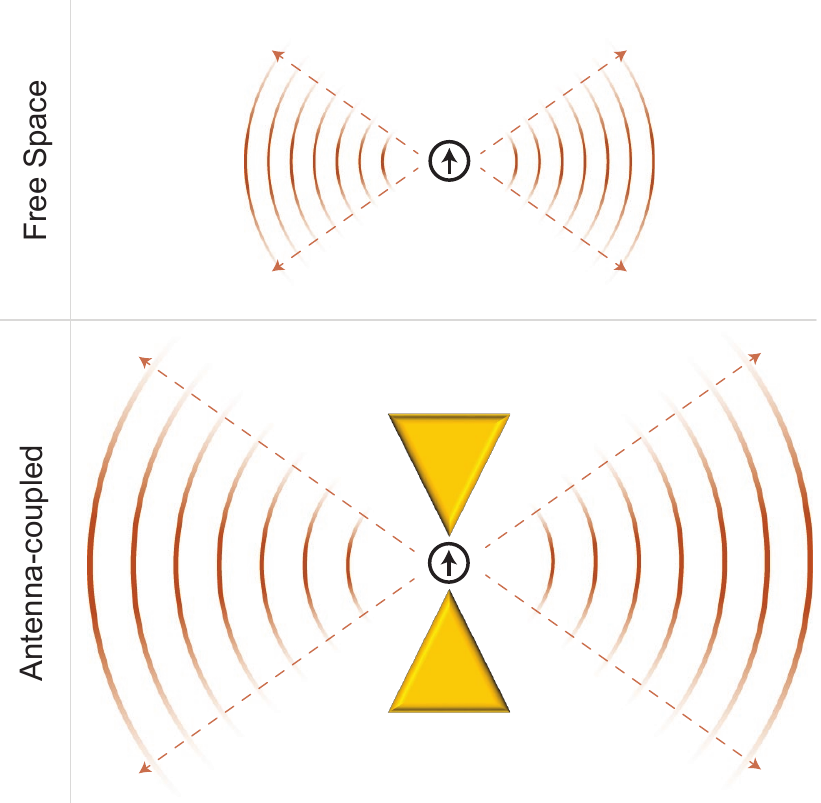}
	\caption{Optical antennas for emission enhancement. A dipolar emitter $\mathbf{p} = \mathbf{n}_{\rm p} \left| \mathbf{p} \right|$ placed at $\mathbf{r}_0$ is radiating in free space (top). Coupling the emitter to an optical antenna leads to an enhancement of radiation (bottom). }
	\label{fig:optant}       
\end{figure}

Consider the situation sketched in Fig.~\ref{fig:optant}. The power emitted by an electric point dipole with dipole moment $\mathbf{p}$ in free space is given as \cite{jackson99}
\begin{equation}\label{eq:po}
P_0 = \frac{\omega^4}{12 \pi \varepsilon_0 c^3} \left|\mathbf{p}\right|^2 ,
\end{equation}
where $\varepsilon_0$ is the vacuum permittivity. This result can be derived in many ways, for our purpose however it is most instructive to express the power emitted by a dipole in terms of the Green's function $\mathbf{G} \left( \mathbf{r}, \mathbf{r}_0 , \omega\right)$. Knowing the Green's function of a given geometry allows us to calculate the electric fields at any location $\mathbf{r}$ for a point dipole $\mathbf{p}$ located at any position $\mathbf{r}_0$ as \cite{novotny06b}
\begin{equation}
\mathbf{E} \left( \mathbf{r} \right) = \frac{1}{\varepsilon_0}\frac{\omega^2}{c^2} \mathbf{G} \left( \mathbf{r}, \mathbf{r}_0 , \omega\right) \cdot \mathbf{p} \, .
\end{equation}

The power dissipated by the dipole is derived from Poynting's theorem and reads as~\cite{novotny06b}
\begin{equation}\label{eq:poynting}
P_\mathrm{tot} = \frac{\omega}{2} \mathrm{Im} \left\{ \mathbf{p}^* \cdot \mathbf{E} \left( \mathbf{r}_0 \right) \right\},
\end{equation}
where $\mathbf{E} \left( \mathbf{r}_0 \right)$ is the field generated by the dipole at its own location $\mathbf{r} = \mathbf{r}_0$. Hence the total power dissipated by a dipole in an arbitrary environment that is described by the Green's function $\mathbf{G} \left( \mathbf{r}, \mathbf{r}_0 , \omega\right)$ is given by
\begin{equation}\label{eq:ptot1}
P_{\rm tot} = \frac{1}{2 \varepsilon_0} \frac{\omega^3}{c^2} \mathrm{Im} \left\{\mathbf{p}^* \cdot \mathbf{G} \left( \mathbf{r}_0, \mathbf{r}_0 , \omega\right) \cdot \mathbf{p}\right\}  .
\end{equation}
For a dipole in free space we use the free-space Green's function $\mathbf{G} = \mathbf{G}_0$ and obtain the result given in equation (\ref{eq:po}), that is, $P_\mathrm{tot}=P_0$. For a dipole oriented along a unit vector $\mathbf{n}_{\rm p}$ as $\mathbf{p} = \mathbf{n}_{\rm p} \left| \mathbf{p} \right|$, equation (\ref{eq:ptot1}) becomes
\begin{equation}\label{eq:ptot2}
P_{\rm tot} = \frac{1}{2 \varepsilon_0} \frac{\omega^3}{c^2} \left|\mathbf{p}\right|^2 \mathrm{Im} \left\{\mathbf{n}_{\rm p} \cdot \mathbf{G} \left( \mathbf{r}_0, \mathbf{r}_0, \omega \right) \cdot \mathbf{n}_{\rm p}\right\}.
\end{equation}

As illustrated in Fig.~\ref{fig:optant}, the power emitted by a point dipole can be enhanced by placing it in the active region of an optical antenna. The total power dissipation however does not only account for the radiated power $P_{\rm r}$ but also for non-radiatively dissipated power $P_{\rm nr}$, which originates from absorption of energy inside the antenna itself. The \textit{antenna radiation efficiency} $\eta_{\rm rad}$ is determined by the ratio of radiative and total power dissipation as
\begin{equation}\label{eq:antennaefficiency}
\eta_{\rm rad} = \frac{P_{\rm r}}{P_{\rm tot}} = \frac{P_{\rm r}}{P_{\rm r} + P_{\rm nr}} \, .
\end{equation}
Hence we find that an optical antenna is able to modify the power radiated by a point dipole as
\begin{equation}
\frac{P_{\rm r}}{P_0} = \frac{P_{\rm tot}}{P_0} \eta_{\rm rad} \, ,
\end{equation}
where $P_0$ and $P_{\rm tot}$ are given by equations (\ref{eq:po}) and (\ref{eq:ptot2}), respectively. It is important to note that emission enhancement is connected to excitation enhancement by reciprocity \cite{bharadwaj09a}.

\subsection{Local density of optical states}\label{sec:ldos}

In many cases, objects that interact with optical antennas are quantum mechanical in their nature, examples being molecules or quantum dots. They are often described as two-level systems with a ground state $\left| g \right>$ and an excited state $\left| e \right>$, the energy difference between them being $\hbar \omega_{eg}$.

An excited two-level system, left to itself in the absence of an antenna, will decay spontaneously and the dynamics of this process is defined by a lifetime $\tau$. The spontaneous decay rate $\Gamma_0 = 1 / \tau$ is given by \cite{loudon83}
\begin{equation}
\Gamma_0 = \frac{\omega_{eg}^3}{3 \pi \varepsilon_0 \hbar c^3} \left|\mathbf{d}_{eg}\right|^2 ,
\end{equation} 
where $\mathbf{d}_{eg} = \left< g \left|\hat{\mathbf{d}}\right| e\right>$ is the transition dipole matrix element. It is defined by the operator $\hat{\mathbf{d}} = -e \hat{\mathbf{r}}$, where $\hat{\mathbf{r}}$ is the position operator. 

It was realized by Purcell in 1946 that the decay rate of a two-level system depends on its environment~\cite{purcell46}. He predicted that---inside a single mode cavity---the decay rate of an emitter tuned to resonance with the cavity is enhanced by a factor of
\begin{equation}\label{eq:purcell}
\frac{\Gamma_{\rm cav}}{\Gamma_0} = \frac{3 \lambda_{eg}^3}{4 \pi^2} \frac{Q}{V} \, ,
\end{equation}
where $\lambda_{eg} = 2 \pi c / \omega_{eg}$, $Q$ is the quality factor of the resonator and $V$ its mode volume. Optical antennas can be viewed as nanoscale resonators, featuring low $Q$-factors but also extremely low mode volumes $V$~\cite{agio12}. Indeed it has been shown that Purcell's factor can be generalized to open and dissipative systems~\cite{sauvan13}. However, it is notoriously difficult to define an appropriate mode volume for optical antennas and can even lead to erroneous results~\cite{koenderink10}.

It turns out that it is not necessary to evaluate the ratio $Q/V$ since this information is contained in the Green's function $\mathbf{G}$. Using the definition of the \textit{partial local density of optical states (LDOS)} \cite{novotny06b}
\begin{equation}\label{eq:ldosGF}
\rho_{\rm p} \left(\mathbf{r}_0, \omega_{ep} \right) = \frac{6 \omega_{ep}}{\pi c^2} \mathrm{Im}\left\{ \mathbf{n}_{\rm p} \cdot \mathbf{G} \left( \mathbf{r}_0, \mathbf{r}_0 , \omega_{eg} \right) \cdot \mathbf{n}_{\rm p}\right\} ,
\end{equation}
the decay rate of a two-level system can be expressed as \cite{novotny06b}
\begin{equation}
\Gamma = \frac{\pi \omega_{eg}}{3 \hbar \varepsilon_0} \left|\mathbf{d}_{eg}\right|^2 \rho_{\rm p} \left( \mathbf{r}_0, \omega_{eg} \right).
\end{equation} 
Hence the decay rate of a two-level system is modified by a factor of 
\begin{equation}\label{eq:ldosenhance}
\frac{\Gamma}{\Gamma_0} = \frac{\pi^2 c^3}{\omega_{eg}^2} \rho_{\rm p} \left( \mathbf{r}_0, \omega_{eg} \right) = \frac{\rho_{\rm p} \left( \mathbf{r}_0, \omega_{eg} \right)}{\rho_0 \left( \omega_{eg}\right)} \, ,
\end{equation}
where $\rho_0 = \omega_{eg}^2 \pi^{-2} c^{-3}$ is the vacuum LDOS. Thus, Purcell's factor (equation (\ref{eq:purcell})) is a special case of equation (\ref{eq:ldosenhance}) \cite{carminati15}.

We conclude that the power dissipation of a classical electric point dipole as well as the decay dynamics of a quantum mechanical two-level system are both dictated by the same quantity, the (partial) LDOS, hence bridging a quantum and classical description. This connection is summarized by the following relation:
\begin{equation}\label{eq:ldos}
\frac{P_{\rm tot}}{P_0} = \frac{\rho_{\rm p}}{\rho_0} = \frac{\Gamma}{\Gamma_0} \, .
\end{equation}

Both the LDOS as well as the Green's function can be derived analytically only for a limited set of geometries. However, equation (\ref{eq:ldos}) allows for the numerical determination of both quantities in arbitrary geometries by calculating power dissipation of a dipole using classical electrodynamics.

%

%

\section{Fundamentals of inelastic electron tunneling}\label{sec:IET}

In conjunction with the experimental investigation of IET, several theoretical models have been developed since its discovery. In the following we will first briefly discuss the most relevant aspects of these models. Next, we will point out the connection between these models and how they are linked to the fundamental aspects of optical antennas discussed previously. Finally we will exemplarily calculate the source efficiency spectrum of IET in the absence of antenna coupling. 

\subsection{Theoretical models}

In the following we will discuss three points of view from which IET is described theoretically. 

\subsubsection{Energy-loss model.} This model is analogous to the energy-loss (EL) of electron beams \cite{ritchie57,garcia10} and was originally proposed and implemented by Davis \cite{davis77}. The primary difference with respect to high energy electron beams is that the source term of the emitted radiation is given by a quantum-mechanical transition current density \cite{davis77}
\begin{equation}\label{eq:qmJ}
\mathbf{J} = \frac{\mathrm{i} e \hbar}{2 m} \left( \psi_\mathrm{R}^* \nabla \psi_\mathrm{L} - \psi_\mathrm{L} \nabla \psi_\mathrm{R}^* \right) ,
\end{equation}
where $\psi_\mathrm{L,R}$ are the electron's eigenfunctions in the left and right electrode of the tunnel junction, respectively. The energy difference between the two states is given by the mode energy $\hbar \omega$. The accompanying transition charge density $\rho$ is determined by $\mathbf{J}$ through the continuity equation, which then allows for the determination of the electromagnetic fields $\mathbf{E}$ from Maxwell's equations. Finally, the inelastic transition rate $\gamma_\mathrm{inel}^\mathrm{EL}$ is calculated as
\begin{equation}
\gamma_\mathrm{inel}^\mathrm{EL} = \frac{-2}{\hbar \omega} \int \mathbf{E}^* \cdot \mathbf{J} \, \mathrm{d}^3 r ,
\end{equation}
i.e. the quantized rate of energy loss. Johannson et al. took an approach similar to Davis, using the same quantum mechanical current density (equation (\ref{eq:qmJ})). They made use of the reciprocity theorem to correlate the radiated electromagnetic field with the electromagnetic field at the position of electron tunneling \cite{johansson90,johansson91,johansson98,aizpurua00}. 

\subsubsection{Current fluctuations (shot noise) model.} The second model describes IET as the result of tunneling current fluctuations (CF) which originate from the quantized nature of the current itself. Here, the rate of IET $\gamma_\mathrm{inel}^\mathrm{CF} $ is related to the power spectrum of these current fluctuations $\left< I_\omega^2 \right>$ as \cite{hone78}
\begin{equation}\label{eq:gammaCF}
\gamma_\mathrm{inel}^\mathrm{CF}  \propto \left< I_\omega^2 \right> =    \left\{
\begin{array}{lll}
\frac{1}{2 \pi R_0} \left( e V_\mathrm{b} - \hbar \omega \right) & \mathrm{for} & \hbar \omega \leq e V_\mathrm{b} \\
0 & \mathrm{for} & \hbar \omega > e V_\mathrm{b}  \\
\end{array} 
\right. \end{equation}
at zero temperature. $R_0$ is the DC resistance of the tunnel junction. Equation (\ref{eq:gammaCF}) is obtained as the Fourier transform of the correlation function of time-varying current fluctuations with the current being evaluated within the transfer-Hamiltonian formalism \cite{bardeen61,bennett68,duke69,rogovin74}. Recently, the power spectrum of current fluctuations was derived using a Landauer-B\"uttiker scattering approach \cite{fevrier18}. Originally derived by Hone et al., equation (\ref{eq:gammaCF}) was used to model IET in a variety of systems \cite{rendell78,laks79,laks80,rendell81,arya83,kirtley83,ushioda86,takeuchi88,szentirmay91,uehara92,bigourdan16}.


\subsubsection{Spontaneous emission model.} The third model describes IET as a spontaneous emission (SE) process, where the transitions occur within the tunnel junction and between electronic states of different energies~\cite{persson92,schimizu92,downes98,uskov16,parzefall17,parzefall18}. Here, the rate of IET $\gamma_\mathrm{inel}^\mathrm{SE}$ is determined by Fermi's golden rule
\begin{equation}\label{eq:gammaSE}
\gamma_\mathrm{inel}^\mathrm{SE} = \frac{2 \pi}{\hbar} \sum_{i, f} \left| \left< \psi_f \left| \hat{H}_\mathrm{int} \right| \psi_i \right> \right|^2 \delta \left( \Delta E - \hbar \omega \right) , 
\end{equation}
where $\hat{H}_\mathrm{int}$ is the Hamiltonian describing the interaction of electronic states with the optical modes of the system, $\psi_{i,f}$ are the initial and final states of the transition process, respectively, and $\Delta E$ is the energy difference between initial and final state, given by the mode energy $\hbar \omega$. $\psi_{i,f}$ are commonly described in the framework of Bardeen's transfer Hamiltonian formalism \cite{bardeen61,bennett68,duke69}. The further implementation varies in that the interaction Hamiltonian is either given by $\hat{H}_\mathrm{int} = - e \hat{\phi}$, where $\hat{\phi}$ is the potential operator of the optical mode \cite{davis77,persson92,schimizu92,downes98}, or by $\hat{H}_\mathrm{int} = - e/m \hat{\mathbf{A}} \cdot \hat{\mathbf{p}}$, where $\hat{\mathbf{A}}$ is the vector potential operator and $\hat{\mathbf{p}}$ is the momentum operator \cite{uskov16,parzefall17,parzefall18}.

\subsection{The link to optical antenna theory}

All three of the previously introduced models have been successfully used to describe experimental results. As the transfer Hamiltonian model~\cite{persson92,schimizu92,downes98,uskov16,parzefall17,parzefall18} is of a perturbative nature its applicability is restricted to strong tunnel barriers where the interaction between the two electrodes is weak. In the high conductivity regime the current fluctuations model \cite{hone78,fevrier18,rendell78,laks79,laks80,rendell81,arya83,kirtley83,ushioda86,takeuchi88,szentirmay91,uehara92,bigourdan16} has proven to accurately describe experimental results. While in most works a single theoretical approach is chosen, it has been shown that the theories of spontaneous emission and finite frequency shot noise are closely related \cite{lu13}. 

In order to understand how the optical properties of a given tunnel junction geometry affect the IET process it is imperative to find a link between the different IET models and the figures of merit that describe an optical antenna. In Sect.~\ref{sec:ldos} we found that it is the (partial) local density of optical states (LDOS) that describes both the modification of the decay rate of a quantum emitter as well as the power dissipation of a localized, oscillating current source. In the energy loss model the electronic properties are determined by the quantum mechanical transition current density, equation (\ref{eq:qmJ}). The power dissipated from this current density, i.e. the rate of IET, depends on the dielectric properties of the environment. As described by equation (\ref{eq:ptot2}), these properties are determined by the Green's function of the geometry---and hence the LDOS (cf. equation (\ref{eq:ldosGF})). 

The proportionality constant in equation (\ref{eq:gammaCF}), the current fluctuations model, is sometimes referred to as the `antenna factor' that depends on the optical properties of the geometry \cite{kirtley83,szentirmay91,dawson84}. More quantitatively, the spectral power emitted by the tunneling device for a given strength of the current fluctuations is determined by the radiation impedance of the geometry \cite{fevrier18}. It can be shown that the radiation impedance, a quantity that is more commonly used in classical antenna theory, is directly linked to the LDOS as well \cite{bharadwaj09a,greffet10}.

The dependence of the inelastic transition rate on the LDOS is most evident in the spontaneous emission model. The direct proportionality of the rate of spontaneous emission to the LDOS in equation (\ref{eq:gammaSE}) is encoded in the sum over final states which includes the number of available optical modes. Starting from equation (\ref{eq:gammaSE}) we arrive at the following expression for $\gamma_\mathrm{inel}^\mathrm{SE}$, the IET rate spectrum~\cite{parzefall17}:
\begin{equation}\label{eq:gammaSE2}
\gamma_\mathrm{inel}^\mathrm{SE} = \frac{\pi e^2}{3 \hbar m^2 \varepsilon_0 \omega} \rho_{\rm p} \sum_{i, f} \left| \left< \psi_f \left| \hat{p} \right| \psi_i \right> \right|^2 \delta \left( \Delta E - \hbar \omega \right) ,
\end{equation}
where $\hat{p}$ is the momentum operator component parallel to the direction of electron tunneling. It is important to note that $\gamma_\mathrm{inel}^\mathrm{SE}$ is of units of Hertz per energy as the total IET rate $\Gamma_{\rm inel}$ is given by $\Gamma_{\rm inel} = \hbar \int_{0}^{\infty} \gamma_\mathrm{inel}^\mathrm{SE} \mathrm{d} \omega$. Equation (\ref{eq:gammaSE2}) effectively disentangles the dependence of the inelastic electron tunneling rate on optical and electronic system properties. The former is encoded in the LDOS ($\rho_{\rm p}$) whereas the latter is described by the momentum matrix element ($\left< \psi_f \left| \hat{p} \right| \psi_i \right>$) and the sum over all initial and final (electronic) states, determined by the \emph{electronic} density of states and the applied voltage~$V_\mathrm{b}$.

\subsection{The source efficiency spectrum of IET}\label{sec:ses}

\begin{figure}
	\centering
	\includegraphics{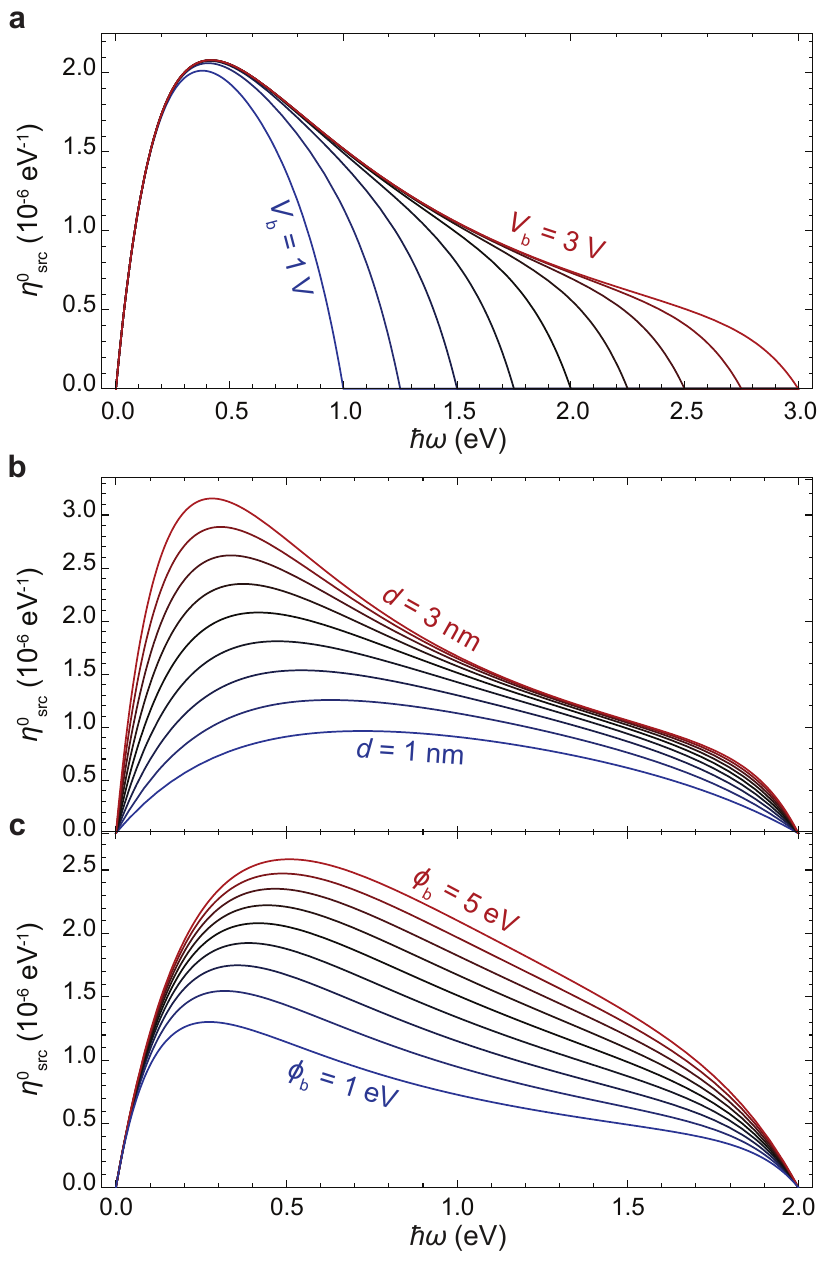}
	\caption{Vacuum source efficiency spectra of IET as a function of (a) applied voltage $V_\mathrm{b}$ for a constant barrier thickness $d=\unit[2]{nm}$ and height $\phi_\mathrm{b} = \unit[3]{eV}$, (b) barrier thickness $d$ for a constant applied voltage $V_\mathrm{b} = \unit[2]{V}$ and barrier height $\phi_\mathrm{b} = \unit[3]{eV}$, (c) barrier height $\phi_\mathrm{b}$ for a constant applied voltage $V_\mathrm{b} = \unit[2]{V}$ and barrier thickness $d=\unit[2]{nm}$. Black curves in (a-c) are identical. Calculations are carried out based on the model described in Appendix~\ref{app:source},  at zero temperature and under the assumption of energy-independent electronic state densities.
%
	}
	\label{fig:theory}       
\end{figure}

The figure of merit of any light source is the external device efficiency $\eta_{\rm ext}$, i.e. the number of photons that are emitted per electron. In most cases, the IET process does not directly result in photon emission but first excites other types of optical modes, e.g. localized and/or propagating SPP modes. The efficiency of this process may be described by a source efficiency $\eta_{\rm src}$. The efficiency with which these optical modes are converted into photons (or another desired output mode) is determined by the radiation efficiency $\eta_{\rm rad}$, as defined in equation (\ref{eq:antennaefficiency}). Hence the external device efficiency is given by 
\begin{equation}\label{eq:etaext}
\eta_{\rm ext} = \eta_{\rm src} \times \eta_{\rm rad} .
\end{equation}

If $\Gamma_{\rm el} \gg \Gamma_{\rm inel}$, $\eta_{\rm src}$ is determined by the ratio of  $\gamma_\mathrm{inel}^\mathrm{SE}$ and the elastic tunneling rate $\Gamma_{\rm el}$. To separate optical and electronic device properties we further define the source efficiency of IET in vacuum as $\eta^0_{\rm src} = \gamma_\mathrm{inel}^\mathrm{SE} (\rho_\mathrm{opt} = \rho_0) / \Gamma_{\rm el}$. We therefore rewrite equation (\ref{eq:etaext}) as
\begin{equation}\label{eq:etaext2}
\eta_{\rm ext} = \eta^0_{\rm src} \frac{\rho_{\rm p}}{\rho_0} \times \eta_{\rm rad} .
\end{equation}

Equation (\ref{eq:etaext2}) illustrates the key parameters determining the external device efficiency of optical antennas driven by IET. While we will go into more detail regarding the optical properties, i.e. the LDOS enhancement  $\rho_{\rm p}/\rho_0$ and the radiation efficiency $\eta_{\rm rad}$, in Sect.~\ref{sec:effrise}, we exemplarily show calculation results for $\eta^0_{\rm src}$ for the case of a rectangular barrier profile in Fig.~\ref{fig:theory}. Details on the model can be found in Appendix~\ref{app:source} and Refs.~\cite{parzefall17,parzefall17b}.

The most important information to be extracted from Fig.~\ref{fig:theory} is that $\eta^0_{\rm src}$ is of the order of $\unit[10^{-6}]{eV^{-1}}$ at optical frequencies. Hence IET is an inherently weak process. Figure~\ref{fig:theory}(a) shows $\eta^0_{\rm src}$ as a function of voltage. It is apparent that the cutoff condition is fulfilled with $\eta^0_{\rm src} = 0$ for $\hbar \omega > e V_\mathrm{b}$. Furthermore, it can be inferred from Fig.~\ref{fig:theory}(b,c) that the overall magnitude and shape of $\eta^0_{\rm src}$ does not strongly depend on the details of the barrier parameters, i.e. its width $d$ and height $\phi_\mathrm{b}$. In general $\eta^0_{\rm src}$ increases with increasing $d$ and $\phi_\mathrm{b}$. This trend can be intuitively understood as follows. A large barrier width increases the interaction time of the tunneling electron with the barrier \cite{buttiker82} and a long interaction time increases the probability of spontaneous emission \cite{uskov16}. On the other hand, a large barrier height favors the inelastic process as it reduces the energy-dependence of the decay constant $\kappa$ that describes the exponential wave function decay inside the tunnel barrier (cf. Appendix~\ref{app:source}). Unfortunately, while increasing $d$ and $\phi_\mathrm{b}$ improves $\eta^0_{\rm src}$, the overall current and emission intensity decrease exponentially.

\section{Optical antennas driven by quantum tunneling: experimental progress}\label{sec:expprog}

%
%

In this section we review recent experiments in antenna-coupled inelastic quantum tunneling. 


\subsection{Efficiencies on the rise}\label{sec:effrise}

%
%
%
%

\begin{figure*}[t]
	\centering
	\includegraphics{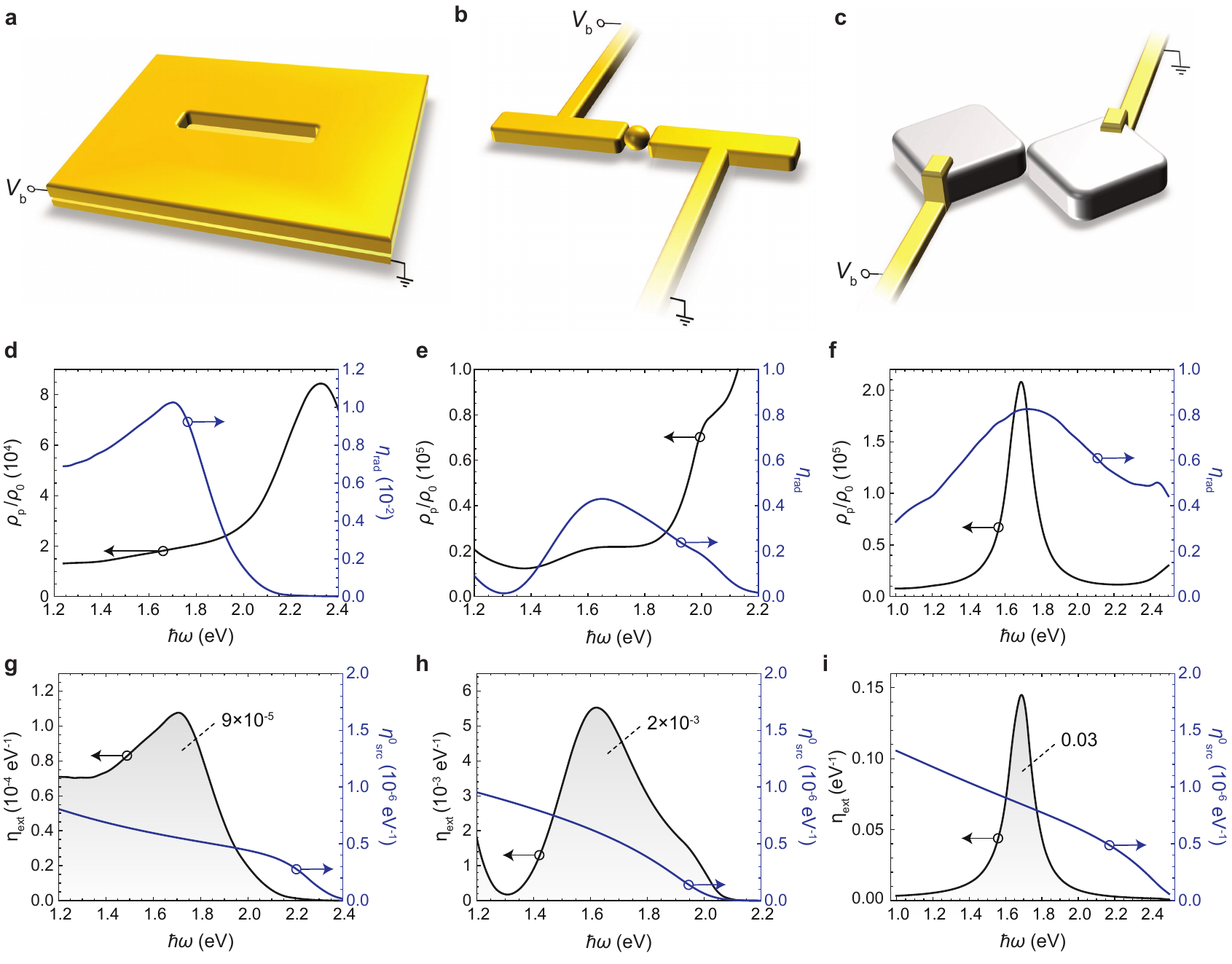}
	\caption{(a-c) Graphical illustrations of (a)~a gold--h-BN--gold vertical MIM structure that is coupled to a slot antenna, cf. Ref.~\cite{parzefall15}, (b)~an electrically connected (gold) dipole antenna with a spherical gold particle placed into its gap, cf. Ref.~\cite{kern15}, (c)~edge-to-edge aligned silver nanocubes, cf. Ref.~\cite{qian18}. (d-f)~Simulated (finite element, COMSOL Multiphysics, wave optics module) LDOS enhancement $\rho_{\rm p}/\rho_0$ (cf. equation (\ref{eq:ldos})) and radiation efficiency $\eta_{\rm rad}$ (cf. equation (\ref{eq:antennaefficiency})) spectra of the antenna geometries shown in panels (a-c), respectively. Dipoles are located (d) at a distance of \unit[50]{nm} from the slot edge, (e)~between the spherical particle and one rod (the other rod is connected to the particle via a conductive channel) and (f)~in the center of the gap between the two silver cubes and oriented parallel to the direction of electron tunneling. (g-i) External device efficiency $\eta_{\rm ext}$ (cf. equation (\ref{eq:etaext2}))and vacuum source efficiency $\eta_{\rm src}^0$ spectra of the antenna geomtries shown in panels (a-c), respectively. The tunnel barrier parameters determining $\eta_{\rm src}^0$ are: (g) $V_\mathrm{b} = \unit[2.3]{V}$, $d=\unit[3]{nm}$, $\phi_\mathrm{b} = \unit[1.5]{eV}$, (g) $V_\mathrm{b} = \unit[2]{V}$, $d=\unit[1.6]{nm}$, $\phi_\mathrm{b} = \unit[2.3]{eV}$, (g) $V_\mathrm{b} = \unit[2.5]{V}$, $d=\unit[1.5]{nm}$, $\phi_\mathrm{b} = \unit[3]{eV}$. The geometrical dimensions of the considered structures are (a,d,g) Slot length: \unit[250]{nm}, slot width/height: \unit[50]{nm}, edge radius: \unit[15]{nm}, h-BN thickness: \unit[3]{nm}, second gold layer thickness: \unit[15]{nm}. (b,e,h) Rod length: \unit[120]{nm}, rod width: \unit[80]{nm}, rod height: \unit[50]{nm}, edge radius: \unit[10]{nm}, sphere radius: \unit[14]{nm}, sphere offset: \unit[20]{nm}, channel diameter: \unit[7.5]{nm}, sphere-rod distance: \unit[1.6]{nm}. (c,f,i) Cube height: \unit[20]{nm}, cube width/length: \unit[70]{nm}, edge radius: \unit[7.5]{nm}, PVP thickness: \unit[1.5]{nm}, cube distance: \unit[1.5]{nm}. The dielectric functions of gold and silver were taken from Refs. \cite{johnson72,mcpeak15}, respectively. The refractive indices of the dielectrics are: glass: $1.52$, h-BN: $1.8$, PVP: $1.52$. For simplicity we omitted the electrical leads to the antenna in the simulations of (b,c).
	}
	\label{fig:comp}       
\end{figure*}

Arguably the most important key issue is the external device efficiency , i.e. the number of photons emitted per tunneling electron (equation (\ref{eq:etaext})). Throughout the initial phase of IET research on metal-insulator-metal (MIM) devices in the 1970s and 80s, efficiencies have remained below $10^{-4}$ \cite{mills82,szentirmay91}, which is not surprising considering the extremely low vacuum source efficiency, cf. Sect.~\ref{sec:ses}. However, it was also claimed, since the very early days
of IET research, that the efficiency with which an electromagnetic MIM mode in the tunnel junction is excited, i.e. the source efficiency $\eta_{\rm src}$, can be of the order of $\unit[10]{\%}$~\cite{davis77}. The high source efficiency is the result of the extreme mode confinement inside a nanoscale MIM gap, leading to LDOS enhancement $\rho_{\rm p} / \rho_0$ values of the order of $10^4$ to $10^5$ \cite{akselrod14,faggiani15,parzefall19}. Unfortunately, the low propagation lengths and severe impedance mismatch with free propagating radiation pose a challenge in achieving a high outcoupling and radiation efficiency $\eta_{\rm rad}$~\cite{parzefall19}. Nonetheless, external quantum efficiencies of $> \unit[1]{\%}$ were recently reported \cite{qian18}. In the following we will analyze three antenna-coupled tunnel junction designs that were experimentally realized. The three designs are illustrated in Fig.~\ref{fig:comp}(a-c).

The first design, illustrated in Fig.~\ref{fig:comp}(a) and realized by Parzefall et al. \cite{parzefall15}, consists of a vertical MIM structure formed by gold electrodes and an insulating hexagonal boron nitride (h-BN) crystal. One of the electrodes is structured into an array of slot antennas. According to equation (\ref{eq:etaext2}), the external device efficiency is determined by three factors, amongst them the LDOS enhancement $\rho_{\rm p}/\rho_0$ and the radiation efficiency $\eta_{\rm rad}$, both optical system properties. Figure~\ref{fig:comp}(d) shows $\rho_{\rm p}/\rho_0$ and $\eta_{\rm rad}$ as a function of energy $\hbar \omega$, as experienced by an electron tunneling through the h-BN crystal in the vicinity of the slot antenna. The LDOS spectrum is to first approximation identical to the LDOS spectrum of the same MIM device in the absence of the antenna \cite{parzefall15}. The radiation efficiency spectrum however reveals that the antenna resonantly enhances the conversion of the MIM mode into photons. In combination with the vacuum source efficiency (cf. Sect.~\ref{sec:ses}) we calculate the external device efficiency spectrum $\eta_{\rm ext}$ of the geometry, shown in Fig.~\ref{fig:comp}(g). The resulting total external quantum efficiency is $9 \times 10^{-5}$, retrieved as the area under the $\eta_{\rm ext}$ spectrum. This result is in satisfactory agreement with the experimentally measured value of $2.5 \times 10^{-5}$ \cite{parzefall15}. Analyzing the individual components it is apparent that the design suffers from a low radiation efficiency which yields an overall low external efficiency of $10^{-5}$ to $10^{-4}$.

With the second design, illustrated in Fig.~\ref{fig:comp}(b), Kern et al. take a different approach \cite{kern15}. Instead of integrating optical antennas with a tunnel junction, they integrate a tunnel junction with an optical antenna. The junction is formed between the two arms of a linear dipole antenna, to which electrical leads have been attached \cite{prangsma12}, and a colloidal gold nanoparticle is placed in between. The nanoparticle forms two tunnel junctions, but one of them is short-circuited according to the authors. The simulated optical properties are displayed in Fig.~\ref{fig:comp}(e). The overall trend of the LDOS spectrum agrees with the LDOS spectrum of panel (d). In addition, as here the junction is an integral part of the antenna, the resonances of the antenna geometry are reflected in the LDOS spectrum with peaks appearing at $\sim \unit[1.6]{eV}$ and $\sim \unit[2.0]{eV}$, in agreement with Ref.~\cite{kern15}. The radiation efficiency of the antenna is significantly higher, reaching a peak value of 0.4. The result of calculating $\eta_{\rm ext}$ from the data presented in Fig.~\ref{fig:comp}(e) is shown in panel (h), resulting in an overall quantum efficiency of $2 \times 10^{-3}$, around one order of magnitude higher than reported. The main origin for this overestimation is most likely differences in the simulated and the actual antenna geometry. We find that the external quantum efficiency strongly depends on the diameter of the conductive channel that connects one arm of the antenna with the particle. The highest efficiency (of the order of \unit[1]{\%}) is reached in the absence of a conductive channel, i.e. for a completely symmetric design. The asymmetry that is caused by the fusion of the particle with one of the two arms appears to adversely affect the external device efficiency.

The third and final design, illustrated in Fig.~\ref{fig:comp}(c) and realized by Qian et al. \cite{qian18}, exhibits the highest experimentally reached device efficiency thus far---more than \unit[1]{\%} of tunneling electrons cause the emission of a photon. The antenna is formed by two chemically synthesized silver nanocubes aligned edge-to-edge. The stabilizing polymer that covers the cubes forms the insulating barrier between them. Contacts are fabricated by focused-ion-beam induced deposition. Our simulation results for this cube-cube geometry are shown in Fig.~\ref{fig:comp}(f). Within the spectral region of interest, the design exhibits a single resonance, reaching LDOS enhancement values of $2\times 10^5$ while exhibiting concomitant radiation efficiencies of the order of 0.8. Our corresponding calculation of the external device efficiency, shown in Fig.~\ref{fig:comp}(i), predicts an overall efficiency of \unit[3]{\%}. The experimentally reported efficiency for the simulated geometry is \unit[1]{\%}. Again we find a satisfactory agreement. The origin of this record efficiency lies in the perfection of colloidally grown metal nanoparticles, the superior optical properties of silver and Qian et al.'s ability to tune the geometrical parameters of the antenna to optimize its optical properties and hence the external quantum efficiency~\cite{qian18}.

\begin{figure*}
	\centering
	\includegraphics{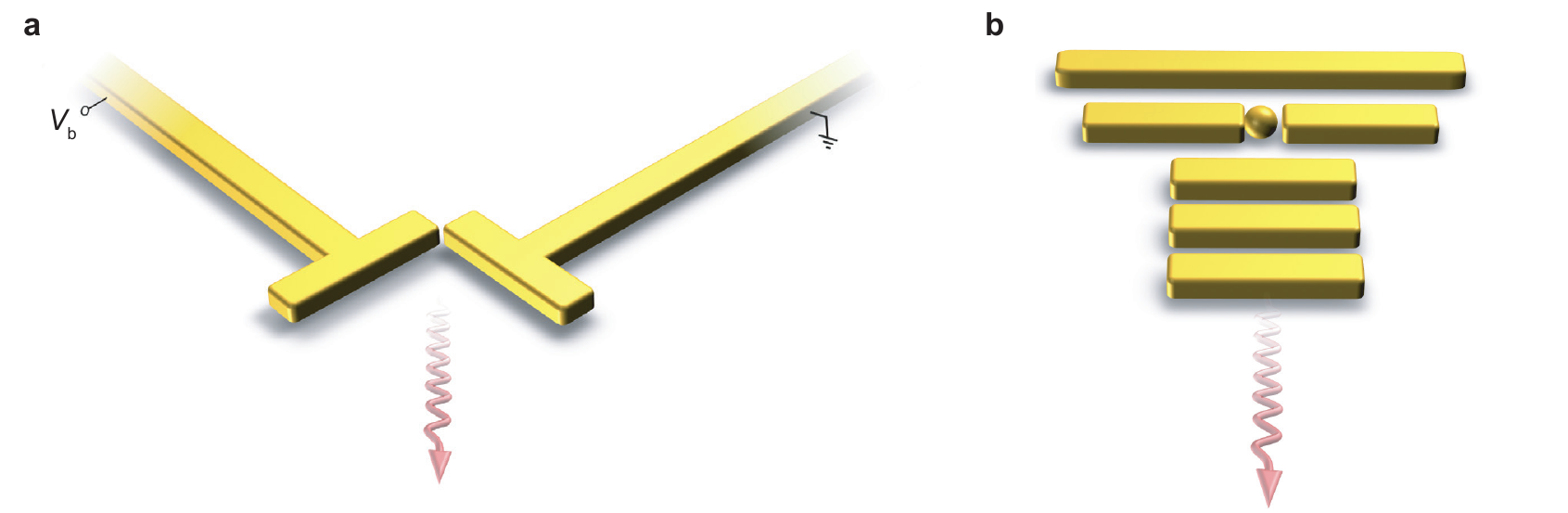}
	\caption{Graphical illustrations of (a) an antenna design realized by Gurunarayanan et al., achieving directivities as high as $\unit[5]{dB}$ \cite{gurunarayanan17} and (b) the design of an electrically driven Yagi-Uda antenna design, proposed by Kullock et al., achieving simulated directivities of $\unit[5.2]{dB}$ \cite{kullock18}.}
	\label{fig:emission}       
\end{figure*}

\subsection{Emission control via antenna design}

One of the key prospects of utilizing optical antennas as sources of light is the ability to control the gain and directivity of radiation. All three designs discussed in Sect.~\ref{sec:effrise} also allow for the deterministic tuning of the emission spectrum by changing the geometrical dimensions of the antenna. In addition, the nature of the antenna resonance was shown to determine the polarization of the emitted radiation \cite{parzefall15,kern15,qian18}.

The directivity of an antenna describes the fraction of power that can be radiated in a certain direction \cite{balanis05}. Directional emission control was previously demonstrated in light-driven optical antennas \cite{taminiau08b,curto10,kosako10,dregely11,dregely14,dasgupta18b}. Directional emission of an optical antenna driven by IET was achieved by Gurunarayanan et al. with the design illustrated in Fig.~\ref{fig:emission}(a) \cite{gurunarayanan17} . It consists of two electrically connected rod antennas that are aligned edge to edge at an angle of $90^\circ$. The directivity of the design originates from the interplay between the dipolar nature of the tunnel junction emission and the quadrupole-like resonance of the combination of the two rod antennas \cite{gurunarayanan17}. Directivities of up to $\unit[5]{dB}$ were demonstrated. 

An alternative directional IET-driven antenna design, based on the well-known Yagi-Uda antenna concept \cite{curto10,kosako10,dregely11,novotny11b}, was proposed by Kullock et al.~\cite{kullock18}, cf. Fig.~\ref{fig:emission}(b). The tunnel junction is based on their original design of a linear dipole antenna (Fig.~\ref{fig:comp}(b)). Simulations suggest that directivities of $\unit[5.2]{dB}$ may be reached by adding suitable reflector and director elements. 

%
%

\subsection{Ultrafast electrooptical signal transduction}

Another prospect of the IET process is its prospective bandwidth. While the speed of an LED is limited by the lifetime of electron-hole pairs, the timescale that is suggested to limit the speed of an IET-driven light source is the tunneling time, a process which happens on femtosecond timescales~\cite{landauer94,shafir12,fevrier18}. Such rapid charge transfer promises unmatched modulation bandwidths. It is important to note however that one also needs to consider the RC time constant $\tau = R C$ as a potential bandwidth limitation, where $R$ is the lead resistance and $C$ the capacitance of the device. In a simple parallel plate capacitor model, $C = \varepsilon A/d$, where $\varepsilon$ is the permittivity of the insulator. We obtain $\tau \sim \unit[10]{ps}$ for exemplary values of $A=\unit[1]{\upmu m^2}$, $d=\unit[1]{nm}$, $\varepsilon = 10 \varepsilon_0$ and $R = \unit[100]{\Omega}$, corresponding to a cutoff frequency of $f_c = 1/(2 \pi \tau)$ of $\sim \unit[15]{GHz}$. 

Experimental tests in this direction have been hindered by the overall low luminance of IET driven light sources. In a proof-of-concept type of experiment, Parzefall et al. demonstrated modulation bandwidths up to \unit[1]{GHz} \cite{parzefall15}. In order to overcome the luminance limitation of their devices, a statistical scheme was employed to demonstrate time-harmonic modulation. The modulated optical signal was sent to two avalanche photodiodes (APDs) via a beam splitter and analyzed by time-correlated single photon counting. This technique, yielding the autocorrelation of the original signal, revealed the modulation of the emitted light. In their initial experiments, the bandwidth was shown to be limited by the timing jitter of the APDs rather than the bandwidth of the device itself \cite{parzefall17}.

Recently, Fevrier and Gabelli used measurements of the light emission from MIM tunneling devices as a probe for the tunneling time \cite{fevrier18}. In conjunction with a current fluctuations model based on the Landauer-B\"uttiker scattering formalism, tunneling times of the order of one femtosecond were inferred. This would suggest an ultimate modulation bandwidth limit of the order of one Petahertz. 

Nonetheless, for the time being the luminance of IET-driven optical antennas limits the practicality of data transduction. In the most recent realization by Qian et al., optical emission powers as high as $\sim \unit[100]{pW}$ at a peak wavelength of \unit[680]{nm} were reached, corresponding to a photon emission rate of $\sim \unit[0.3]{GHz}$ from a single antenna. 

%

\subsection{Novel applications}

Going beyond the shear optimization of the device efficiency, few groups have devoted their research towards the exploration of new applications of IET-driven devices. We will discuss two such reports in the following.

Dathe et al. employed inelastic electron tunneling for the excitation of SPPs in a metal-insulator-semiconductor geometry \cite{dathe16}. For the unmodified, planar devices, they observe weak and spectrally broad light emission. They further placed colloidal gold nanoparticles on top of their devices, covered with adsorbents of different size. As the scattering of SPPs by the gold nanoparticles spectrally varies as a function of the distance between the gold nanoparticle and the gold film, a shift as a function of adsorbent size was detected. As this `nanoruler' is electrically excited, it does not require an external illumination source.

In an entirely different geometry, Wang et al. utilized IET to monitor chemical reactions \cite{wang18}. Their geometry consists of vertical arrays of gold nanorods which form one electrode of a metal-air-metal junction geometry, with the other electrode being formed by liquid eutectic gallium indium. In this configuration, signatures of the nanorod array resonance are observed in photon emission spectra. Subsequently, the air tunnel gap is replaced with a reactive polymer. Variations in emission intensity as a function of gas-composition were observed, attributed to hot-electron activation of oxidation and reduction in the tunnel junction.

\section{Outlook: future directions and unexplored territories}\label{sec:outlook}

%


\subsection{Efficiency limitations and bypass strategies}

In Sect.~\ref{sec:effrise} we have seen that efficiencies have been improved by several orders of magnitude within the first few iterations of antenna-coupled tunnel junction designs. While this trend seems very promising, can we expect it to continue? 

Returning to equation (\ref{eq:etaext2}) we may assess the improvement potential of the parameters that determine $\eta_{\rm ext}$. Figure \ref{fig:comp} suggests that there is not much room for radiation efficiency improvement. The LDOS enhancement  inside an MIM gap scales with $d^{-3}$ for nanometer scale gaps, derived from classical electrodynamics \cite{faggiani15}. To first approximation, the vacuum source efficiency of IET scales with $d^2$ \cite{parzefall18}. These dependencies combined yield a $d^{-1}$ dependence of $\eta_\mathrm{ext}$, suggesting a divergence of the efficiency with vanishing gap size. However, the LDOS enhancement is ultimately limited by quantum mechanics, i.e. non-locality as well as the charge exchange due to tunneling itself, limiting the maximum LDOS enhancement achievable \cite{ciraci12,savage12a,zhu16b}. Additionally, as the voltage required to excite a certain mode of interest has to be larger than its mode energy, the reduction in tunnel barrier thickness is limited by the risk of electrostatic breakdown which occurs in the range of $\unit[1]{V/nm}$ for most insulators. These factors suggest that efficiencies are ultimately limited. Nonetheless, other prospects of IET-driven devices such as its bandwidth, footprint and spectral tunability may outweigh the efficiency limitations depending on the application. 

Moreover, the realizations studied so far are almost exclusively based on MIM tunnel gaps. As the electronic properties of metals are generally very similar, all realizations feature qualitatively similar vacuum source efficiencies (cf. Sect.~\ref{sec:ses}). As is well known within the STM light emission community, the efficiency of IET is affected by the energy-dependent variation of the electronic density of final states \cite{nazin03,hoffmann04,berndt93c,lutz13,chen14,kuhnke17}. This dependence may be explored in future devices to favor the inelastic process over the elastic process by bandstructure engineering. Alternatively, efficiencies may be increased by orders of magnitude in devices that support resonant inelastic electron tunneling \cite{belenov87,uskov16}. Promising platforms for the realization of such strategies are molecular tunnel junctions \cite{tao06,du16} and van der Waals heterostructures \cite{geim13,novoselov16,parzefall19}. Several theoretical works have predicted the efficient excitation of graphene plasmons by IET in graphene-based tunneling devices, promising an excitation source for the far-infrared and Terahertz regime \cite{svintsov16,enaldiev17,vega17}.

\subsection{Designer photon statistics}

\begin{figure}
	\centering
	\includegraphics{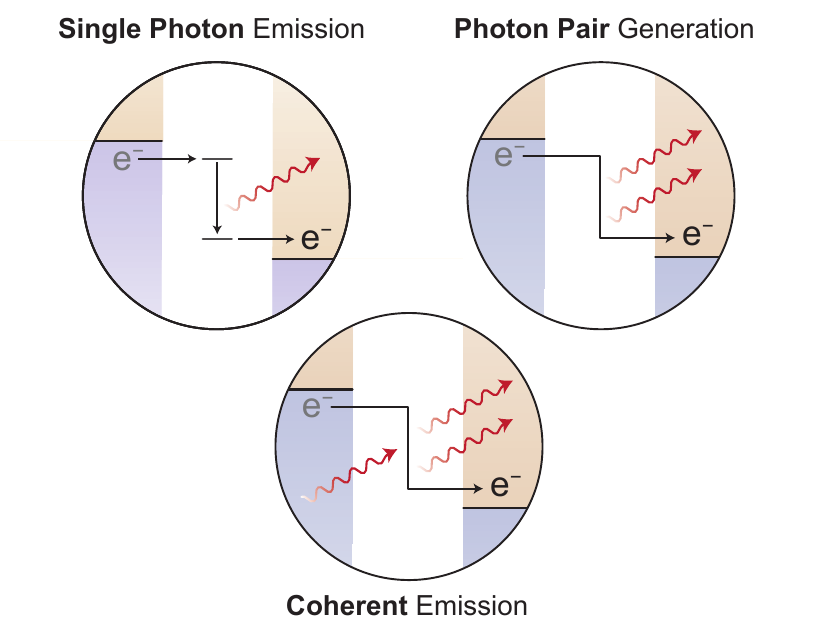}
	\caption{Suggested mechanisms and tunnel junction configurations that yield modified photon emission statistics.}
	\label{fig:statistics}       
\end{figure}

The photon statistics of light emitted from solid state tunneling devices has not been paid much attention to so far. However, reports from STM light emission experiments as well as theoretical works suggest that IET-driven devices may be designed to exhibit different types of photon statistics as schematically depicted in Fig.~\ref{fig:statistics}.

\subsubsection{Single photon emission.} Merino et al. observed anti-bunched photon statistics when analyzing light emission from C$_{60}$ films on silver, locally excited using an STM \cite{merino15}. The emission behavior was attributed to the injection of charge carriers into localized structural defects. Furthermore, electrically driven single-photon emission from zinc-phtalocyanine molecules on top of a sodium chloride-covered silver substrate was demonstrated \cite{zhang17,luo19}. The light is generated by locally injecting charge carriers into the molecule with an STM. 
Taking inspiration from these reports, localized states supported by a variety solid state systems \cite{aharonovich16} may be utilized to realize single photon emission from an antenna-coupled tunnel junction.

\subsubsection{Photon pair generation.} Initial STM studies of the emitted photon statistics reported bunching of the emission which was however attributed to system properties not related to the tunnel junction and the nature of the emission process itself \cite{silly00b,perronet06}. To exclude external influences, Leon et al. carried out time-correlated single photon counting experiments of the light emitted from a metal-vacuum-metal junction (STM) at low temperatures (\unit[4]{K}) and ultrahigh vacuum ($\unit[10^{-11}]{mbar}$) \cite{leon18}. A significantly increased probability for simultaneous (bunched) photon emission was found with the second order autocorrelation function reaching values as high as 17, suggesting the emission of photon pairs as illustrated in the corresponding pictogram in Fig.~\ref{fig:statistics}.

\subsubsection{Coherent emission.} Describing IET as a spontaneous emission process naturally implies the possibility for stimulated emission. 
Stimulated inelastic electron tunneling was theoretically discussed as a possible explanation for observations of negative differential resistance in MIM point contact diodes  \cite{siu76,drury80}. While these theoretical considerations have not materialized thus far, its implications are profound, suggesting the possibility of an electrically driven, nanoscale, coherent light source.

\section{Conclusions}

We have discussed the fundamentals of optical antennas and inelastic electron tunneling. Regarding the former we have pointed out the ability of optical antennas to modify the radiative behavior of classical and quantum mechanical entities via their modification of the local density of optical states. Regarding the latter we have discussed several theoretical frameworks that describe the IET process and their respective links to optical antenna theory. Moreover we have revisited the concept of viewing IET as an excitation source with a  source spectrum that is determined by the electronic properties of the tunnel junction and shaped by its environment through the local density of optical states.

In light of these foundations we discussed the recent experimental demonstrations of antenna-coupled tunnel junctions. Additionally we discussed some of the prospects of IET-driven optical antennas such as the ability to control emission spectrum, polarization and directivity by varying the antenna dimensions and arrangements. 

Finally we pointed out strategies that may lead to further improvements in device efficiencies and novel device functionalities.

\section*{Acknowledgments}

This work was supported by the Swiss National Science Foundation (grant no. 200021\_165841). 

\appendix

\section{IET source efficiency in vacuum}\label{app:source}

In the following we derive the equations for the IET source efficiency in vacuum $\eta_{\rm src}^0$ (cf. Sect.~\ref{sec:ses}) for a one-dimensional, rectangular barrier model. $\eta_{\rm src}^0$ is defined as 
\begin{equation}\label{eq:etasrc}
\eta_{\rm src}^0 = \frac{\gamma_\mathrm{inel}^0}{\Gamma_{\rm el}},
\end{equation}
the ratio of the spectral inelastic tunneling rate in vacuum $\gamma_\mathrm{inel}^0$ and the elastic tunneling rate $\Gamma_{\rm el}$. In the framework of Bardeen's transfer Hamiltonian formalism \cite{bardeen61,bennett68,duke69}, both are calculated as perturbative interactions between the electronic wave functions of the (unperturbed) electrodes. At zero temperature the elastic tunneling rate reads as~\cite{chen93}
\begin{equation}\label{eq:gammael}
\Gamma_{\rm el} = \frac{2 \pi}{\hbar} \int \limits_{0}^{e V_{\rm b}} \left| t  \right|^2  \rho_{\rm L} (E) \, \rho_{\rm R} (E) \, {\rm d} E ,
\end{equation}
whereas the spectral inelastic tunneling rate, derived from Fermi's golden rule, is given by~\cite{parzefall17}
\begin{equation}\label{eq:gammainel}
\gamma_{\rm inel}^0  =  \frac{\pi e^2}{3\hbar \omega m^2 \varepsilon_0} \rho_0  \int \limits_{\hbar \omega}^{eV_{\rm b}} \left| p  \right|^2  \rho_{\rm L} (E) \, \rho_{\rm R} (E - \hbar \omega) \, {\rm d} E.
\end{equation}
$t$ and $p$ are the matrix elements for elastic and inelastic tunneling, respectively, $\rho_{\rm L/R}$ are the electronic densities of states of the left and right electrode and $m$ is the electron mass. Inserting equations (\ref{eq:gammael}) and (\ref{eq:gammainel}) into equation (\ref{eq:etasrc}) yields the following expression for $\eta_{\rm src}^0$:
\begin{equation}
\eta_{\rm src}^0 = \frac{e^2}{6 m^2 \varepsilon_0 \omega} \rho_0 \frac{\int_{\hbar \omega}^{eV_{\rm b}} \left| p  \right|^2 \rho_{\rm L} (E) \, \rho_{\rm R} (E - \hbar \omega)  \, \mathrm{d} E}{\int_0^{e V_{\rm b}} \left| t \right|^2 \rho_{\rm L} (E) \, \rho_{\rm R} (E) \, \mathrm{d} E} .\label{eq:deveffA}
\end{equation}
The matrix elements are given by \cite{bardeen61,reittu95}
\begin{equation}\label{eq:matel1}
	t (E)  =  \frac{\hbar^2}{2 m} \left[ \psi_{\rm L} \frac{{\rm d} \psi_{\rm R}^*}{{\rm d} z} - \psi_{\rm R}^* \frac{{\rm d} \psi_{\rm L}}{{\rm d} z} \right]_{z = z_0}
\end{equation}	and
\begin{equation}\label{eq:matel2}
	p (E, \hbar \omega) =  - \mathrm{i} \hbar \int \limits_{0}^{d} \psi_{\rm R}^* (E-\hbar \omega) \frac{{\rm d}}{{\rm d} z} \psi_{\rm L} (E) \, {\rm d} z .
\end{equation}
The matrix elements and consequently the tunneling rates as well as the spectral efficiency depend on the modeling of the electronic wave functions $\psi_\mathrm{L/R} (z)$ in the barrier adjacent to the left and right electrode, respectively. The unperturbed (in the limit $d \rightarrow \infty$) wave functions for a rectangular barrier of height $\phi$ and thickness $d$ are given by
\begin{align}
\psi_\mathrm{L} (z) = \frac{2 k_\mathrm{L}}{k_\mathrm{L} + {\rm i} \kappa_\mathrm{L}} e^{-\kappa_\mathrm{L} (z+d/2)}, \quad & z \geq -d/2  \label{eq:psi1} \\ 
\psi_\mathrm{R} (z) = \frac{2 k_\mathrm{R}}{k_\mathrm{R} + {\rm i} \kappa_\mathrm{R}} e^{-\kappa_\mathrm{R} (d/2 - z)}, \quad & z \leq d/2 \label{eq:psi2}
\end{align}
where $k_\mathrm{L/R} = [2 m E_\mathrm{L/R}/\hbar^2]^{1/2}$ and $\kappa_\mathrm{L/R} = [2 m (E_\mathrm{F} + \phi - E_\mathrm{L/R})/\hbar^2]^{1/2}$ are propagation and decay constants, respectively, with $E_\mathrm{F}$ being the Fermi energy. Evaluating equations (\ref{eq:matel1}) and (\ref{eq:matel2}) based on (\ref{eq:psi1}) and (\ref{eq:psi2}) yields
\begin{align}
t  = &   \frac{\hbar^2}{2 m} \left[ \psi_\mathrm{L} \frac{{\rm d} \psi_\mathrm{R}^*}{{\rm d} z} - \psi_\mathrm{R}^* \frac{{\rm d} \psi_\mathrm{L}}{{\rm d} z} \right]_{z = z_0} \nonumber
\\  = & \frac{\hbar^2}{m} \frac{4 \kappa k^2 }{k^2+\kappa^2} e^{-\kappa d}  ,
\label{eq:t}
\end{align}
with $\kappa_\mathrm{L} = \kappa_\mathrm{R} = \kappa$ and $k_\mathrm{L} = k_\mathrm{R} = k$, and 
\begin{align}
p  = & \left< \psi_\mathrm{R} \left| \hat{p}_z \right| \psi_\mathrm{L} \right> \nonumber \\  = & - {\rm i} \hbar \int_{-d/2}^{d/2} \psi_\mathrm{R}^* \frac{\mathrm{d}}{\mathrm{d} z} \psi_\mathrm{L} \, \mathrm{d} z \nonumber\\  = &  \frac{8 {\rm i} \hbar \kappa_\mathrm{L} k_\mathrm{L} k_\mathrm{R}}{(k_\mathrm{L} + {\rm i} \kappa_\mathrm{L})(k_\mathrm{R} - {\rm i} \kappa_\mathrm{R})}e^{-\frac{d}{2}(\kappa_\mathrm{L} + \kappa_\mathrm{R})}\frac{\sinh \left(\frac{d}{2}(\kappa_\mathrm{L} - \kappa_\mathrm{R})\right)}{\kappa_\mathrm{L} - \kappa_\mathrm{R}} .  \label{eq:p}
\end{align}

\end{document}